\documentclass [8 pt,twocolumn]{article}

\usepackage{amsmath,amsthm,amscd,amssymb}

\usepackage[small, centerlast, it]{caption}
\usepackage{epsfig}
\usepackage[titles]{tocloft}
\usepackage[latin1]{inputenc}
\usepackage{verbatim} %for comments via \begin{comment}
\usepackage[active]{srcltx} %inverse search
\usepackage{fancyhdr}
\usepackage{caption}

\def\b1{{1\!\!1}}

\def\sH{{\mathsf H}}

\def\bC{{\mathbb C}}           %%%  complex numbers and so on

\def\bI{{\mathbb I}}

\def\bB{{\mathbb B}}
\def\bN{{\mathbb N}}

\def\bR{{\mathbb R}}

\def\bZ{{\mathbb Z}}

\def\gP{{\mathfrak P}}

\def\beq{\begin{eqnarray}}
\def\eeq{\end{eqnarray}}

\usepackage{sectsty}
\sectionfont{\normalsize}%\normalfont
\subsectionfont{\normalsize\normalfont\itshape}
\newtheoremstyle{thm}
{12pt}% space above
{12pt}% space below
{\itshape}% body font
{}% h indent amount
{\itshape\bfseries}% theorem head font
{}% punctuation after theorem head
{1em}% space after theorem head
{}% theorem head spec (can be left empty, meaning `normal')

\theoremstyle{thm}

\title{\Large Open-loop quantum control as a resource for secure communications}
\author{ Davide Pastorello\\\normalsize Department of  Mathematics, University of Trento,\\
\normalsize INFN-TIFPA\\
 \normalsize via Sommarive 14, 38123 Povo (Trento), Italy.}

\date{}

\begin{document}

%\maketitle

\twocolumn[\begin{@twocolumnfalse}

\maketitle
\begin{abstract}
\noindent
Properties of unitary time evolution of quantum systems can be applied to define quantum cryptographic protocols. Dynamics of a qubit can be exploited as a data encryption/decryption procedure by means of timed measurements, implementation of an open-loop control scheme over a qubit increases robustness of a protocol employing this principle.
\end{abstract}

\vspace{1. cm}	

\end{@twocolumnfalse}]

\section{Introduction}

One of the most prominent practical applications of \emph{quantum information theory} is \emph{quantum cryptography} \cite{n,m,K}, in particular the so-called Quantum Key Distribution (QKD) where a transmission of quantum information (e.g. through polarized photons in an optical fiber) is used to create a shared key between two clients. In classical cryptography, if Alice wants to communicate a secret message $x\in\mathbb B^N$, where $\mathbb B=\{0,1\}$ and $N\in\bN$, over a public communication channel, she can adopt the \emph{one-time pad} security: She randomly generates a second bit string $y\in\mathbb B^N$ (the key) and sends $x+y$ to Bob. The receiver who knows the string $y$ can decrypt the message simply adding $y$, the communication is perfectly secure only if the key $y$ is securely exchanged and manteined secret.
\\Quantum information processes can be applied to distribute a secret key in order to use it as a one-time pad. 
 Information exchange in a quantum channel prevents eavesdropping attacks exploiting principles of Quantum Mechanics: An eavesdropper (Eve) cannot clone an unknown quantum state (\emph{no-cloning theorem}, see appendix A) thus she can gain information only performing measurements on qubits and quantum effects of these measurements can be detected showing that used communication channel is not secure.
\\
Mathematical formulation of quantum mechanics can be derive from a set of postulates that we briefly recall in order to introduce principles of quantum mechanics and basic features of quantum systems that are relevant in the present discussion, in particular we refer to \emph{finite-dimensional case}: 
\\
\\
i) A complex vector space $\sH$ with inner product $\langle\,\,\,|\,\,\,\rangle$ (a Hilbert space) is associated with any quantum system.  
%ii) Physical observables related to the considered quantum system are described by selfadjoint operators on $\sH$.
\\
ii) Physical (pure\footnote{There exists a more general notion of quantum state, the \emph{mixed state}, that is not necessary in our discussion}) states of the considered quantum system are equivalence classes of unit vectors $\psi\in\sH$ where $\psi\sim\psi'$ iff $\psi=e^{i\theta}\psi'$ for some $\theta\in\bR$. A unit vector $\psi\in\sH$ representing a quantum state is called \emph{state vector}\footnote{We adopt \emph{Dirac formalism}: A unit vector of the Hilbert space is denoted by the ket $|\psi\rangle$, a vector of dual space is denoted by the bra $\langle\psi|$, the inner product of two vectors is denoted by $\langle\psi|\phi\rangle$, the outer product by $|\psi\rangle\langle\phi|$. }.
\\
iii) Time evolution of an isolated system is described by a continuous one-parameter group of unitary operators $\{U(t)\}_{t\in\bR^+}$ acting on state vectors. If $|\psi_1\rangle$ is the state of the system at time $t_1$ and $|\psi_2\rangle$ is the state of the system at time $t_2>t_1$ then:
\beq
|\psi_2\rangle=U(t_2-t_1)|\psi_1\rangle.
\eeq
iv) A measurement process on a quantum system is described by a collection of positive operators $\{E_k\}$ satisfying  $\sum_k E_k=\bI_\sH$ called \emph{positive operator-valued measure} (POVM), the index $k$ runs in the set of all possible outcomes of the measurement, so it is a real number\footnote{There is a more general notion of POVM that also describes measurements with a non-discrete set of outcomes \cite{mo}, it is defined as a measure on the Borel $\sigma$-algebra of $\bR$.}. The probability to measure $k$ when the system is in the state $|\psi\rangle$ is:
 \beq
p_\psi(k)=\langle \psi|E_k\psi\rangle.
\eeq 
%After the measurement (with outcome $n$) the state of the system is:
%\beq
%\psi'= \frac{E_n\psi}{\sqrt{\langle\psi|E_n\psi\rangle}}.
%\eeq 
v) If a quantum system is composed by two subsystems $A$ and $B$ that are respectively described in
Hilbert spaces $\sH_A$ and $\sH_B$, then the total system is described in the Hilbert space given by the Hilbert space tensor product
$\sH_A\otimes \sH_B$.
\\
\\
A special class of POVMs is that of \emph{projective measurements} whose elements are orthogonal projectors $P_k$, i.e. operators on $\sH$ satisfying $P_k^*=P_k$ and $P_k^2=P_k$. In this case if the measurement is performed when the state of the system is $|\psi\rangle$ and it produces the outcome $k$ then the state of the system after the measurement is: 
\beq
|\psi'\rangle= \frac{P_k|\psi\rangle}{\sqrt{\langle\psi|P_k\psi\rangle}},
\eeq
and a selfadjoint operator on $\sH$, called \emph{observable}, can be defined:
\beq
A=\sum_k kP_k,
\eeq
so the possible outcomes of the measurement described by the POVM $\{P_k\}$ are the eigenvalues of $A$.  In particular the selfadjoint operator $H$ describing total energy of the system, called \emph{Hamiltonian}, is the \emph{generator} of the group $\{U(t)\}_{t\in\bR^+}$, that is the unique (by Stone theorem \cite{mo}) self-adjoint operator such that $U_t=\exp[-iHt]$ for any $t\in\bR^+$.
\\
\\
A \emph{classical bit} is the basic unit of information and it can be realized by a physical system which admits only two states, namely $0$ and $1$. While a \emph{quantum bit} is a two-level quantum system (i.e. it is described in a two-dimensional Hilbert space) then its general state $|\psi\rangle$ is given by a superposition of two orthogonormal state vectors that can be denoted by $|0\rangle$ and $|1\rangle$:
\beq\label{psi}
|\psi\rangle=\alpha |0\rangle +\beta |1\rangle\qquad \alpha,\beta\in\bC,
\eeq
thus the Hilbert space of such quantum system is isomorphic to $\bC^2$. Since state vector $|\psi\rangle$ has unit norm, i.e. $|\langle\psi|\psi\rangle|^2=1$, then $|\alpha|^2+|\beta|^2=1$ and the state vector admits the following angular representation:
\beq\label{bloch}
|\psi\rangle=\cos\frac{\theta}{2}|0\rangle+e^{i\varphi}\sin\frac{\theta}{2}|1\rangle,
\eeq
for $\theta,\varphi\in\bR$, so the state of a qubit can be representd on the so-called \emph{Bloch sphere}, according to (\ref{bloch}) the states $|0\rangle$ and $|1\rangle$ corresponds to the poles. Postulate ii) reads that unit vectors differing by a multiplicative phase factor are physically indistinguishable then (pure) states can be described by rank-1 orthogonal projectors $|\psi\rangle\langle\psi|$ in a one-to-one correspondence.
\\
In order to give a physical realization of a quibit one can consider polarization states of a photon, spin states of an electron, an atom oscillating between ground state and a single excited state. Considering the example of the electron, we denote the spin down state as $|0\rangle$ and the spin up state as $|1\rangle$. The measurement process of spin (e.g. realized in a Stern-Gerlach apparatus) is described by the projectors:
\beq
P_0=|0\rangle\langle 0|\quad \mbox{and}\quad P_1=|1\rangle\langle 1|.
\eeq
If the state of the electron is the linear superposition (\ref{psi}) then the probability to measure \emph{spin down} (value 0) is $p(0)=\langle\psi|0\rangle\langle0|\psi\rangle=|\alpha|^2$ and the probability to measure \emph{spin up} (value 1) is $p(1)=\langle\psi|1\rangle\langle1|\psi\rangle=|\beta|^2$. After the measurement the state of the electron is $|0\rangle$ if the outcome of measurement is $0$, while after the measurement the state is $|1\rangle$ if the measured value is 1. Generally speaking one can map states $|0\rangle$ and $|1\rangle$ into classical bit states 0 and 1 during a measurement.

%Spin measurement is described by the projectors:
%\beq
%M_0=|0\rangle\langle 0|\quad \mbox{and}\quad M_1=|1\rangle\langle 1|.
%\eeq
%For example, if the qubit state is $|+\rangle$, the probability to measure 0 is $p(0)=\langle +|M_0^*M_0 +\rangle=\langle+|0\rangle\langle0|+\rangle=\frac {1}{2}$ and the state after the measurement is $|0\rangle$. The measurement of the same observable (spin, polarization,...) can be performed w.r.t. another basis, e.g. $\{|+\rangle,|-\rangle\}$, so $|+\rangle$ maps into 0 and $|-\rangle$ into 1 and the measurement operators are $M_0=|+\rangle\langle +|$ and $M_1=|-\rangle\langle -|$.

\section{Basic control scheme of a qubit}
\noindent
The control scheme discussed in this section is \emph{open-loop}, this means the control law is designed and completely determined before the experiment and then used without modification during the experiment \cite{DD}.
\\
Consider a two-level quantum system (qubit) described by the Hamiltonian:
\beq\label{H}
H=H_0+u_1(t)H_1(t)+u_2(t)H_2(t)\quad t\in[0,T],
\eeq
where $H_0$ represents the internal energy of the system, $u_1=u_1(t)$ and $u_2=u_2(t)$ are control functions defined on the real interval $[0,T]$, $H_1$ and $H_2$ are  time-dependent Hamiltonians describing interaction of the system with external fields.
%\beq
%F_1(t)=A_1(t)\cos(\omega_1t+\phi_1),\\
%F_2(t)=A_2(t)\cos(\omega_2t+\phi_2),
%\eeq
%where $A_1$ and $A_2$ are the pulse envelopes and the frequencies $\omega_1$ and $\omega_2$  are off-resonant w.r.t. transition frequency $\omega_0$ then the external field $F_i$ induces an oscillation between $|0\rangle$ and $|1\rangle$ with frequency $\tilde\omega=\sqrt{(\omega_i-\omega_0)^2 +\omega_r}$, where $\omega_r$ is the Rabi frequency.  
\\
The control function $u_1$ and $u_2$ are defined piecewise as follows: Let $\{[t_{n-1},t_n]\}_{n=1,...,N}$, with $0=t_0<t_1<\cdots<t_N=T$, be a partition of $[0,T]$ into $N\in \bN$ subintervals, for any $n=1,...,N$ the control functions are given by:
\beq\label{controls}
u_1(t):=b_n\in\bB=\{0,1\}\quad\forall t\in[t_{n-1},t_n],\\\label{controls1}
u_2(t):=1-u_1(t).\qquad\qquad\qquad\qquad\quad\quad
\eeq
Thus $u_1$ and $u_2$ are piecewise constant function completely determined by a vector in $\bB^N$ which defines a time sequence of non-overlapping control pulses.
This structure also implies that the control of the system is attuated turning on and off the external fields according to a suitable time sequence (bang-bang control).
\\
Time evolution of the controlled system is described by the solution $U=U(t)$ of the operatorial Schr\"odinger equation:
\beq\label{se}
i\hbar \frac{d}{dt}U(t)=\left[ H_0+\sum_{i=1,2}u_i(t)H_i(t)\right]U(t),
\eeq
with initial condition $U(0)=\bI_2$. If the state of the system is $|\psi_0\rangle$ at time $t=0$ then at time $t=T$ the system is in the final state $|\psi_f\rangle=U(T)|\psi_0\rangle$. Hence a certain time evolution (and in particular a certain final state) can be achieved\footnote{In general not all states can be achieved from an initial state in a finite time and a \emph{reachable set} can be defined.} acting on the controls $u_1,u_2$, i.e. choosing a suitable sequence $(b_1,...,b_N)\in\bB^N$ of control pulses.
\\
The control sequence to achieve the desired evolution corresponds to a factorization of the unitary operator $U(T)$. Adopting the interaction picture, solution of (\ref{se}) is $U(t)=U_0(t)U_I(t)$ where $U_0(t)=\exp[-i/\hbar H_0 t]$ is the evolution operator of the free system. The interaction component $U_I$ satisfies the equation:
\beq\label{ie}
i\hbar \frac{d}{dt} U_I(t)=U_0^*(t)\left[\sum_{i=1,2}u_i(t)H_i(t)\right]U_0(t)U_I(t),
\eeq
as the direct calculation shows (see appendix B).  $U_I(t)$, with  $t\in[t_{n-1},t_n]$, can be factorize as follows:
\beq
U_I(t)=V_n(t)U_I(t_{n-1}),
\eeq
where the operator $V_n$ is obatined integrating equation (\ref{ie}):
$$
V_n(t)=\exp\!\!\left[-\frac{i}{\hbar}\!\int_{t_{n-1}}^t \!\!\!\!\!\!\!U_0^*(\tau)\left(\sum_{i=1,2}u_i(\tau)H_i(\tau)\right)\!\!U_0(\tau) d\tau\right]\!\!=$$
$$\quad\,\,=\exp\left[-\frac{i}{\hbar}\int_{t_{n-1}}^t\!\!\!\!\!\! U_0^*(\tau)\left(b_nH_1(\tau)+\right.\right.\qquad\qquad\qquad
$$
\beq
\qquad\qquad\qquad\qquad\left.\left.+(1-b_n)H_2(\tau)\right)U_0(\tau) d\tau\right].\quad\,\,
\eeq
Iterating on the total interval this elementary factorization we can write the decomposition of the desired unitary operator $U(T)$:
\beq
U(T)=U_0(T)U_I(T)=\exp\left[-\frac{i}{\hbar}H_0T\right]V_NV_{N-1}\cdots V_1
\eeq
where
$$
V_n=\exp\left[-\frac{i}{\hbar}\int_{t_{n-1}}^{t_n}\!\!\!\!\!\! U_0^*(t)\left(b_nH_1(t)+\right.\right.
$$
\beq
\qquad\qquad\qquad\qquad\left.\left.+(1-b_n)H_2(t)\right)U_0(t) dt\right].\quad\,\,
\eeq
Generally speaking, factorization of a unitary operator into a product of generators of dynamical Lie group determines the sequence of control pulses to achieve a target evolution, then it is a strategy of constructive quantum control, as widely explained in \cite{s}.   
\\
In the following we will be interested to drive back the system from the final state $|\psi_f\rangle$ to the initial one $|\psi_0\rangle$, this procedure is mathematically described by the application of the adjoint evolution operator $U^*(T)$ to the final state, this is given by:
\beq\label{8}
U^*(T)=U_I^*(T)U_0^*(T)=V_1^*\cdots V_N^*\exp\left[\frac{i}{\hbar}H_0T\right],
\eeq
where
$$
V^*_n=V_n^{-1}=\exp\left[\frac{i}{\hbar}\int_{t_{n-1}}^t\!\!\!\!\!\! U_0^*(t)\left(b_nH_1(t)+\right.\right.\qquad\qquad\quad
$$
\beq\label{9}
\quad\qquad\qquad\qquad\qquad\left.\left.+(1-b_n)H_2(t)\right)U_0(t) dt\right].\quad\,\,
\eeq
If a qubit state is decomposed in the basis of eigenvectors $|0\rangle$ and $|1\rangle$ of $H_0$ (let $E_0$ and $E_1$ be the corresponding eigenvalues), $|\psi\rangle=\alpha|0\rangle+\beta|1\rangle$ then the action of operator $U_0(T)$ (or $U_0^*(T))$ only produces relative phases:
\beq
\exp\left[\pm\frac{i}{\hbar}H_0T\right]|\psi\rangle= \alpha e^{\pm\frac{i}{\hbar}E_0T}|0\rangle+\beta e^{\pm\frac{i}{\hbar}E_1T}|1\rangle,
\eeq
which do not affect the probabilities $|\alpha|^2$ and $|\beta|^2$ to measure values $E_0$ and $E_1$ respectively. We say that $|\psi\rangle$ is equal to $U_0(T)|\psi\rangle$ up to relative phases implying the two states are indistinguishable by means of a measurement process described by projectors $P_0=|0\rangle\langle 0|$ and $P_1=|1\rangle\langle1|$.
\\
Thus from (\ref{8}) and (\ref{9}) it is clear that in order to re-obtain the initial state (up to relative phases due to drift term $H_0$) of the system from $|\psi_f\rangle$ in a time interval $[\tau,\tau+T]$ one can switch the sign of the control fields
and set the control sequence $(b'_1,...,b'_N)$ defined as $b'_n:=b_{N+1-n}$ for every $n=1,...,N$. The $(b'_1,...,b'_N)$-controlled dynamics with sign-switched interaction terms is described by the evolution operator $U_0(t)U_I^*(t)$ that does not correspond to the adjoint of $U(t)=U_0(t)U_I(t)$. However from the point of view of the projective measurement $\{P_0,P_1\}$ these two evolution operators are indistinguishable as clarified by the example below.
\\
\\
\textbf{Remark}: In view of further applications, let us discuss the following example: Let $b\in\bB^N$ be a bit string describing a control sequence for the Hamiltonian (\ref{H}) and $U(t)=\exp[-\frac{i}{\hbar} Ht]$ is the associated evolution operator which can be factorized as $U(t)=U_0(t)U_I(t)$ where $U_I$ satisfies equation (\ref{ie}).  Consider the initial state $|0\rangle$ (eigenstate of free Hamiltonian $H_0$) at time $t_0=0$ and the evolved state $|\psi\rangle=U(t)|0\rangle$ for some $t\geq 0$. Swhitching the sign of control fields and choosing the control sequence defined by $b'\in\bB^N$, the system can be driven back from the pure state $|\psi\rangle\langle\psi|$ to the pure state $|0\rangle\langle0|$ in a time interval of lenght $t$, as shown by:
\beq\label{id}
U_0(t)U_I^*(t)\,|\psi\rangle\langle\psi|\,U_I(t)U_0^*(t)=|0\rangle\langle0|,
\eeq
(See appendix C for a proof of (\ref{id})).
If the initial state is not an eigenstate of free Hamiltonian then the action of $U_0(t)U_I^*(t)$ on the final state  does not produce the initial state but another pure state that is indistinguishable from the initial one by means of a measurement described by $|0\rangle\langle0|$ and $|1\rangle\langle1|$.
\\
\\
The model of a qubit controlled by a series of control pulses in the bang-bang scenario described above admits an immediate physical realization in terms of a half-spin particle interacting with an external magnetic field. In this case control pulses are actually implemented in time, on the other hand if the qubit is realized by a photon the control pulses are implemented in space rather than in time by means of passive optical elements. Regarding applicability, photons are an effective solution for long-distance quantum communications \cite{m}. Therefore let us discuss an analogue model that is convenient for physical realizations with photons.
\\
Consider a two-level quantum system described by the Hamiltonian with time-independent coupling terms:
\beq\label{hh}
H=\sigma_z+u_1(t)\sigma_x+u_2(t)\sigma_y,
\eeq
where $\sigma_{x,y,z}$ are the \emph{Pauli matrices}:
\beq
\sigma_x=\left(\begin{array}{ll}
0 & 1 \\
1 & 0
\end{array}
\right),\quad
\sigma_y=\left(\begin{array}{ll}
0 & -i \\
i & 0
\end{array}
\right),\quad
\eeq
$$
\sigma_z=\left(\begin{array}{ll}
1 & 0 \\
0 & -1
\end{array}
\right),\qquad\qquad
$$
and the control functions $u_1$ and $u_2$ are defined as before in (\ref{controls}) and (\ref{controls1}). Thus the final state after the evolution governed by (\ref{hh}) in a \emph{partitioned} time interval $[0,T]$ is given applying to the initial state the unitary operator:
\beq\label{factor1}
U(T)=e^{-i\sigma_zT}V_1\cdots V_N,
\eeq
where:
\beq\label{factor2}
V_n=\exp[-i (b_n\sigma_x+(b_n-1)\sigma_y)\Delta t  ].
\eeq
and $\Delta t=t_n-t_{n-1}$. Thus the open-loop control law of the qubit is individuated again by a bit-string $(b_1,...,b_n)\in\bB^N$ like in the above analysis.
\\
A possible physical realization of the quantum system with Hamiltonian (\ref{hh}) is a half-spin particle in a magnetic field with time varying components in the $x$ and $y$ directions, while the free motion is the \emph{Larmor precession} around $z$-axis. A control sequence $b_n$ determines which component of the magnetic field is active within time interval $[t_{n-1},t_n]$ then it corresponds to a sequence of rotations on the Bloch sphere around $x$ and $y$ axes, as shown in (\ref{factor1}) and (\ref{factor2}), whose angles depends on the amplitude of intervals $\{[t_{n-1},t_n]\}_{n=1,...,N}$.
\\
If the physical system is a linearly polarized photon circulating in a ring cavity then $|0\rangle$ and $|1\rangle$ represent the states of horizontal and vertical polarization respectively. The control pulses are implemented in space by means of passive elements on the photon path. The very short time of flight of the photon inside the optical elements implies that one of these elements can be modeled as a control pulse with a very narrow envelope. Operators $\exp[-i\varphi \sigma_x]$ and $\exp[-i\varphi \sigma_y]$ with $\varphi\in[0,2\pi]$ can be experimentally implemented by a suitable oriented \emph{dispersive wave retarder} where $\varphi$ is the phase accumulated by the photon after a single passage through the optical element \cite{fly}. Then the operator $V_n$ defined in (\ref{factor2}) can be realized putting a dispersive element on the photon path during the time interval $[t_{n-1},t_n]$, this corresponds to an integer number $q$ of cycles in the optical cavity producing an angle $q\varphi$:
\beq
V_n=\left\{\begin{array}{ll}
\exp [-i (q \varphi) \sigma_x] \quad\mbox{for}\quad b_n=1\\
\exp [-i (q \varphi) \sigma_y] \quad\mbox{for}\quad b_n=0\,
\end{array}
\right. .
\eeq
Indeed $V_n$ does not describe the action of a single control pulse applied in $[t_{n-1},t_n]$ but a series of $q$ very narrow-size control pulses. 
In order to obtain the inverse dynamics one has to consider the adjoint of above operator that is given by:
\beq
V_n^*=V_n^{-1}=\left\{\begin{array}{ll}
\sigma_y V_n\sigma_y  \quad\mbox{for}\quad b_n=1\\
\sigma_z V_n\sigma_z  \quad\mbox{for}\quad b_n=0\,
\end{array}
\right. ,
\eeq
and then it can be realized by means of optical elements (wave-plates) acting on polarization states as Pauli matrices (that are the simplest quantum gates).
\\
Free dynamics inside the cavity corresponds to a phase-noise along $z$-axis produced by cavity mirrors, in a triangular cavity (three cavity mirrors) the polarization transformation during a time interval $[0,T]$ is:
\beq
U_0(T)=\exp[-ip(3\Phi)\sigma_z],
\eeq
where $\Phi$ is the relative phase due to a single mirror and $p$ is the number of round trips in the interval $[0,T]$. Therefore we recover the factorized operator (\ref{factor1}) in the photon setting.

\vspace{0.cm}
\section{Dynamical protocol}

In this section we present some ideas to implement secure transmission of data or a cryptographic key over a public channel exploiting the properties of unitary time evolution of quantum systems. 
\\
We can classify quantum cryptographic protocols in two main categories: \emph{Preparation/measurement protocols} (like BB84 \cite{bb}) and \emph{entanglement-based protocols} (like E91 \cite{e}). Let us propose a third kind of protocols which we call \emph{dynamical protocols}, where data encryption/decryption procedures are realized by means of dynamics of a quantum system. Assume two parties (Alice and Bob) have choosen a computational basis $\{|0\rangle,|1\rangle\}$ fixing two orthonormal states of a quantum system described in a bi-dimensional Hilbert space. Suppose Alice wants to send the  classical bit $0$ to Bob, she does not transmit the qubit in the state $|0\rangle$ but she can transmit the qubit in the evolved state $|\psi\rangle_t=U(t)|0\rangle$, where $U(t)=\exp[-iHt]$ is the evolution operator induced by the Hamiltonian $H$ ($\hbar=1$). Therefore quantum dynamics gives rise to an encryption of datum to be transmitted. The secret datum can be decrypted applying $U^*(t)$ which represents another time evolution of the qubit after quantum trasmission, as illustrated in the next section where the above control scheme will be applied. 
\\
Let us discuss the simplest architecture based on the notion of \emph{quantum evolution as encryption} which can be realized with a non-controlled isolated system. Suppose a qubit is confined in an experimental apparatus called Alice's slot (it can be a photon circulating in a ring cavity or a half-spin particle in a magnetic trap) and it is described by the time-independent Hamiltonian $H$. There is also a Bob's apparatus that is a copy of Alice's slot such that the quibit can be trasnmitted from Alice to Bob over a quantum channel that is assumed to be \emph{noiseless}, i.e. quantum states are transmitted unaltered. Assume $|0\rangle$ and $|1\rangle$ are the eigenstates of $H$ and the computational basis for the information processing is given by the states $|+\rangle=\frac{1}{\sqrt 2}(|0\rangle+|1\rangle)$ and $|-\rangle=\frac{1}{\sqrt 2}(|0\rangle-|1\rangle)$ representing classical bits $1$ and $0$ respectively, if the qubit is prepared in $|+\rangle$ at time $t=0$ then its dynamics is described by:
\beq
|\psi\rangle_t=e^{-iHt }|+\rangle=\frac{1}{\sqrt 2}\left( e^{-iE_0t}|0\rangle+e^{-i E_1 t}|1\rangle\right)
\eeq
for $t>0$, then the probability to measure $1$ during the evolution is:
\beq\label{pb}
\gP_1(t)=|\langle +|\psi\rangle_t|^2=\frac{1}{4}\left | e^{iE_0t}+e^{iE_1t}\right|^2.
\eeq
The non-negative roots of (\ref{pb}) are $\tau_n=\frac{(2n+1)\pi}{E_1-E_0}$ with $n\in\bZ^+$ then $\gP_0(\tau_n)=1$, where $\gP_0$ is the probability to measure the value $0$.
\\
 A dynamical protocol can be summarized as follows:
\\
\\
\textbf{Step 1} (Preparation).
If Alice wants to send the classical bit $0$ (represneted by $|-\rangle$) to Bob, she prepares the qubit in the evolved state $|\psi\rangle_\tau=\exp[-iH\tau]|+\rangle$ obtaining an encrypted datum. 
%\\
%\textbf{Step 1}. Using a public classical channel Alice declares the time $T_0$ at which she transmits the qubit in $|\psi\rangle_\tau$ over a quanum channel. $T_0$ is the receiving time in Bob's slot.
\\
\textbf{Step 2} (Quantum communication). Alice transmits the qubit in $|\psi\rangle_\tau$ over a quantum channel.
\\
\textbf{Step 3} (Classical communication). Over a classical public channel Alice declares the time $T_1$ at which Bob must perform the measurement on the received qubit.
\\
\\
After quantum transmission the qubit evolves in the Bob's slot according to evolution operator $U(t)=\exp[-i H t]$. The measurement time $t=T_1$ is choosen by Alice such that the state of qubit when Bob performs the measurement is $|\psi\rangle_{\tau_n}$ which differs from $|-\rangle$ by a phase factor, i.e. the probability to measure $0$ on $|\psi\rangle_{\tau_n}$ is 1. In other words if the quibit is received by Bob at time $T_0$ then Alice selects $T_1$ so that $\tau+(T_1-T_0)=\tau_n$. Otherwise if Alice wants to transmit the classical bit $1$ then she declares a measurement time $T_1'$ so that $\tau+(T'_1-T_0)=\tau'_n$ where $\{\tau'_n\}_{n\in\bZ}$ are time values such that $\gP_1(\tau'_n)=1$, i.e. the non-negative roots of $\gP_0(t)=|\langle -|\psi\rangle_t|^2$.
\\
Thus the state $|\psi\rangle_\tau$ can be used to encrypt both classical bits $0$ and $1$. Only the value of $T_1$ in the classical communication discriminates if decryption procedure in Bob's slot gives $0$ or $1$.    
\\
About unconditional security of such simple protocol, since an unknown quantum state is impossible to clone (see appendix A)  the only way to decrypt the quantum information is implementing time evolution of the qubit according to $H$ and performing the measurement at time communicated by the sender. An eavesdropper (Eve) can intercept the quibit over the quantum channel, however she must have an exacy copy of Alice and Bob's slots to state the right time evolution of the system. In this case she can intercept the classical data $T_1$ and decrypt the information, nevertheless she must re-prepare the qubit in the encrypted state and re-send it to the receiver in order to hide her attack. Such eavesdropping attack produces a delay on the quantum transmission which can be detected by Bob. In fact if a classical authentication protocol is implemented over the classical channel in order to ensure that the right person is at the end of the line then Bob performs the measure at time $T_1$ discovering no qubit arrives in his slot. Hence an eavesdropping produces a lack of information. 
\\
If Eve does not re-send qubit in the original state but she decides to provide a \emph{fake qubit} to the receiver then she produces a randomized datum, i.e. the attack is an error source. During the transmission of a message or QKD, errors can be detected appending mutually agreed bit sequences to the message, otherwise an error estimation and reconciliation procedure can be adopted on the shared key \cite{m}.

\vspace{0.cm}

\section{Controlled dynamical protocol}

Despite theoretical simplicity, the general protocol described in the previous section is not convenient in physical realizations, with trapped half-spin particles for instance, because of high frequencies and susceptibility to noise of free dynamics. In this section a basic open-loop quantum control scheme is applied to define a dynamical protocol in order to obtain a significant robustness w.r.t. the general structure described above.
\\In this section we want to define a protocol of quantum communication (which can be applied also for a QKD) exploiting the controlled dynamics of a qubit. The main idea is setting up a quantum communication where a qubit in an arbitrary superposition state $\alpha|0\rangle+\beta|1\rangle$ is sent representing an encrypted information, before quantum transmission the sender tells to the receiver how to control qubit dynamics to achieve $|0\rangle$ or $|1\rangle$ in a selected time interval. After controlled evolution (decryption) the receiver can perform a measurement.
%An eavesdropper who wants to gain information from such communication must know exactly what are the interaction term in the total Hamiltonian of the system. If she has an apparatus for applying the right external control field she can intercept the classical information about controls then she can intercept the qubit over the quantum channel, apply the suitable evolution operator decrypting the information. However in order to maintain herself hidden she must resend the qubit in the original state to the receiver with an inevitable delay in the transmission that can be detected.
\\
Let us describe how our protocol works: The communication system is made by Alice's slot where the qubit is described by the Hamiltonian (\ref{H}) and the Bob's slot where the interaction term in the Hamiltonian is sign-switched. The slots are equipped with synchronized clocks and a default time interval $[0,T]$ partitioned in $N$ subintervals $\{[t_{i-1},t_i]\}_{i=1,...,N}$ is \emph{a priori} fixed.
\\
The scheme below shows the steps of the protocol:
\\
\\
\textbf{Step 1} (Encryption). Alice prepares a qubit in the state $|0\rangle$ (or $|1\rangle$) e.g. performing a projective measurement, this is the information she wants to send.
Then she randomly generate a bit-string in $\bB^N$ representing a control sequence $(b_1,...,b_N)$. She sets the controlled evolution for an arbitrary time interval $[0,T']$ with $T'=t_m<t_N=T$ obtaining the encrypted datum $|\psi\rangle=U(T')|0\rangle$.
\\
Note the sub-string $(b_1,...,b_m)\in\bB^m$ (i.e. the values of control function $u=u(t)$ for $t<t_m$) is sufficient to drive the system from $|0\rangle$ to $|\psi\rangle$ in the time $T'$.
\\
\textbf{Step 2} (Classical communication): Using a classical communication channel Alice sends to Bob the 
pair $(T_0,b')\in \bR\times\bB^N$ where $b'\in\bB^N$ is the suitable control sequence for Bob defined as follows:
$$b'_i:=b_{m+1-i}\qquad\mbox{for} \quad1\leq i\leq m,$$
$$b_i':=\mbox{Rand}(\bB)\qquad \mbox{for}\quad m<i\leq N.$$
The bits $b_i$, for $i>m$, are randomly taken because are not relevant to implement the unitary operation required for decryption. $T_0$ is the receiving time of quantum transmission, i.e. \emph{when} the qubit arrives in Bob's slot and the controlled evolution according to $b'$ can start.
\\
\textbf{Step 3} (Quantum communication): Alice sends the qubit in the state $|\psi\rangle$ to Bob over the quantum channel according to timing information sent in the previous step. Then at time $T_0$ controlled evolution of the qubit starts in Bob's slot.
\\
\textbf{Step 4} (Decryption): Alice tells Bob \emph{when} performing a measurement on qubit. Datum $T_1$ is transmitted over a classical channel.
Bob follows the instructions received on classical channel  performing a measuremnt when the qubit state is $U_0(T_1-T_0)U_I^*(T_1-T_0)|\psi\rangle=U_0(T')U_I^*(T')|\psi\rangle$ that correspond to $|0\rangle$ up to a multiplicative phase factor.
\\
\\
Let us remark why the encryption procedure has not stated exploiting the whole interval $[0,T]$ to control system dynamics and why preparing quibit in the state $|\psi\rangle=U(T)|0\rangle$ is a naive strategy. In this case, an eavesdropper (Eve) could explicitely obtain the operator $U(T)$ from  the classical message of Step 1 solving equation (\ref{ie}). Thus she would know that the qubit is in the state $|\psi\rangle=U(T)|0\rangle$ or $|\varphi\rangle=U(T)|1\rangle$, she would be able to perform a measurement w.r.t. the new orthonormal basis $\{|\psi\rangle,|\varphi\rangle\}$ gaining information and re-sending the qubit to Bob in the original state, giving place to a perfectly hidden eavesdropping. For this reason a time sub-interval is adopted, however fixing total interval is crucial to well-define a control sequence.
\\
An eavesdropper, with an exact copy of Alice and Bob's slots, can attack the communication process with a woman-in-the-middle strategy playing the Bob's part for Alice and the Alice's part for Bob. To avoid this fact a classical authentication protocol must be implemented over the classical channel in order to ensure that the right person is at the end of the line \cite{m}. Classical authentication is a typical device of QKD protocols adopting also classical communications like celebrated BB84 \cite{bb} and E91 \cite{e}.
\\
If Eve intercepts the classical message gaining control sequence $b'\in\bB^N$ and receiving time $T_0$, she can intercept the qubit implementing the controlled evolution at time $T_0^*$ ( $T_0^*\not = T_0$ is the right receiving time for Eve depending on the receiving time $T_0$ for Bob, let us suppose Eve knows $\Delta T=T_0-T_0^*$ because of her deep knowledge of the quantum channel). Then she intercepts the second classical communication gaining $T_1$ so she performs the measurement at time $T_1^*=T_1-\Delta T$ completing decryption phase in her copy of Bob's slot. Since she cannot copy an unknown quantum state the only way to re-transmit the original qubit state to Bob is implementig the whole decryption phase and re-prepare the qubit. Then Bob will receive the qubit with a time delay $\tau>T'$. 
The presence of delay in quantum transmission and consequent lack of information proves an eavesdropping attack occured. 
\\
If Eve wants to prevent a time delay the only thing she can do is sending a qubit in a new arbitrary state to Bob who would receive a quibit at expected time $T_0$ but this produces a random bit (error) which can be detected appending mutually agreed bit sequences to the message, otherwise an error estimation procedure can be adopted on the shared key after a QKD with our protocol. Every eavesdropping without delay produces randomized incoming qubits in Bob's slot, i.e. it is always an error source\footnote{While in BB84 an eavesdropping attack produces randomized qubits only if Eve chooses a different measurement basis w.r.t. to Alice and Bob \cite{bb} otherwise transmission is unperturbed.}. 
\\
Let us summarize a general eavesdropping attack assuming Eve has an exact copy of Bob's slot and a synchrnonized clock with clients' clocks:
\\
\\
\textbf{Stage 1}
{Eve intercepts the classical message, gaining instructions about decryption: Initial time $T_0$ of controlled evolution and control sequence $b\in\bB^N$.}
\\
\textbf{Stage 2}
She moves an intercept-and-resend attack over the quantum channel. She implements decryption procedure on the intercepted qubit performing a measurement at time $T_1$ declared by Alice in the second classical communication. She gains secret information sent by Alice. In the third stage Eve must take a decision between two possible strategies.
\\
\textbf{Stage 3(a)}
She reprepares the qubit in the original state and resends it to Bob. Time duration of this procedure implies the qubit is not in Bob's slot when he performs the measurement at time $T_1$. Thus Eve produces a time delay corresponding to an information disappearance.
\\
\textbf{Stage 3(b)}
{During decryption phase in Eve's slot, she provides a fake qubit to Bob according to expected receiving time in order to hide her presence. Bob's implements decryption and performs the measurement obtaining a completely randomized outcome.}
\\
\\
An effective eavesdropping attack can achieve the secret information however it certainly produces indelible marks revealing Eve's presence. More precisely if the information is intercepted then there is no way to resend it to Bob in order to maintain an eavesdropper hidden. For this reason the presented protocol is particularly effective for quantum key distribution. If Eve adopts the startegy of Stage 3(b) during QKD then Alice's key and Bob's key do not match because they are completely scorrelated by Eve's fake qubits. 
\\
We can slightly generalized the protocol from transmission of a single qubit to transmission of a qubit-string. Suppose Alice wants to send a qubit $k$-string in the state: 
\beq\label{qstring}
|0\rangle\otimes|0\rangle\otimes|1\rangle\otimes|0\rangle\otimes|0\rangle\otimes|1\rangle\otimes|0\rangle\otimes%|1\rangle\otimes|1\rangle\otimes|1\rangle\otimes|0\rangle\otimes|1\rangle\otimes|1\rangle\otimes|0\rangle\otimes
\cdots
\eeq
the encryption is implemented controlling the evolution of qubits in a selected time interval $[0,T']$, i.e. applying $U^{\otimes k}(T')$ to (\ref{qstring}) obtaining the encrypted string:
\beq\label
||\psi\rangle\otimes|\psi\rangle\otimes|\phi\rangle\otimes|\psi\rangle\otimes|\psi\rangle\otimes|\phi\rangle\otimes|\psi\rangle\otimes%|\phi\rangle\otimes|\phi\rangle\otimes|\phi\rangle\otimes|\psi\rangle\otimes|\phi\rangle\otimes|\phi\rangle\otimes|\psi\rangle\otimes
\cdots
\eeq
where $|\psi\rangle=U(T')|0\rangle$ and $|\phi\rangle=U(T')|1\rangle$. The quantum transmission is characterized by $\Delta t$ that is the time interval between the transmissions of a single qubit and the following one in the string. Bob receives control sequence and $T_0$  over the classical channel. Bob will activate the control sequence at time $T_0$ and he will start to measure single qubits at time $T_1$ repeating measurements at $T_1+l\Delta t$ for $l=1,...,k-1$. Hence a qubit-string with arbitrary lenght can be encrypted with the same amount of classical information.

\vspace{0. cm}

\section{Conclusions}

In the present work a general open-loop scheme to control a single qubit is discussed with some hints about physical realizations. Then the scheme with two control functions is proposed to define a quantum cryptographic protocol  where controlled dynamics of qubit (or a string of qubits) gives rise to an encryption procedure and the values of controls are transmitted in a classical communication. In particular the control law is encoded in a bit-string, called \emph{control sequence}, that contains a redundant information. Decryption can be implemented by the receiver once known the control sequence and the time at which he must perform a measurement on the received qubit. Unconditional security is guaranteed by the fact that the unique way to intercept information is implementing a controlled time evolution of the qubit for decryption causing a detectable delay in transmission.\\       
 Generally speaking, an abstract dynamical protocol based on the free dynamics of a qubit can be physically unfeasible, otherwise an open-loop controlled time evolution of a qubit can be designed to increase the feasibility and robustness of such a protocol.

\vspace{0 cm}

\section*{Appendix A. No-cloning theorem}

Cloning an unknown quantum state is not possible in general \cite{m, no-cloning}. Suppose the existence of a \emph{cloning machine}, i.e. a composite quantum system which is described in the Hilbert space given by the tensor product $\sH\otimes \sH$ such that there is a unitary operator (a time evolution of the total system) which allows to duplicate the state of one subsystem. More precisely the initial state of the composite system is $|\psi\rangle\otimes |\Psi_i\rangle$, the copying procedure is given by the action of the unitary operator $U$:
\beq
U(|\psi\rangle\otimes|\Psi_i\rangle)=|\psi\rangle\otimes|\psi\rangle,
\eeq
for any $|\psi\rangle$. A no-go theorem on the existence of $U$ can be easily proved: Consider two initial state $|\psi\rangle, |\phi\rangle\in\sH$ to be cloned. The copying procedure is:
$$U(|\psi\rangle\otimes|\Psi_i\rangle)=|\psi\rangle\otimes|\psi\rangle,$$
$$U(|\phi\rangle\otimes|\Psi_i\rangle)=|\phi\rangle\otimes|\phi\rangle,$$
taking the inner product of above terms we have: $\langle\psi|\phi\rangle=(\langle\psi|\phi\rangle)^2$. This equation holds if and only if $|\psi\rangle=|\phi\rangle$ or $\langle\psi|\phi\rangle=0$. Thus for a pair of general quantum states there is not a copying procedure (a unitary operator $U$) and a general cloning device cannot exist. 
\\
This result has a remarkable impact on quantum information, in particular on security of quantum channels, in fact it implies that Eve cannot gain information from unknown qubits but she must perform measurements corrupting information.

\section*{Appendix B. Derivation of equation \ref{ie}}

Starting from equation (\ref{se}):

$$i\hbar\frac{d}{dt} U_0(t)U_I(t)=\left[ H_0+\sum_{i=1,2}u_i(t)H_i(t)\right]\!\!U_0(t)U_I(t),\,$$
$$i\hbar\left(\frac{d}{dt} U_0(t)U_I(t)+U_0(t)\frac{d}{dt}U_I(t)\right)=\qquad\qquad\qquad\quad$$
$$\qquad\qquad\qquad=\left[ H_0+\sum_{i=1,2}u_i(t)H_i(t)\right]U_0(t)U_I(t),$$
\\
$$H_0U_0(t)U_I(t)+i\hbar U_0(t)\frac{d}{dt}U_I(t)=\qquad\qquad\qquad\qquad$$
$$\quad\quad=H_0U_0(t)U_I(t)+\left[\sum_{i=1,2}u_i(t)H_i(t)\right]U_0(t)U_I(t),$$
\\
$$i\hbar U_0(t)\frac{d}{dt}U_I(t)=\left[\sum_{i=1,2}u_i(t)H_i(t)\right]U_0(t)U_I(t),\qquad$$
\\
$$i\hbar\frac{d}{dt}U_I(t)=U_0^*(t)\left[\sum_{i=1,2}u_i(t)H_i(t)\right]U_0(t)U_I(t),$$
\\
obtaining equation (\ref{ie}).

\section*{Appendix C. Proof of identity \ref{id}}

Given the orthonormal basis $\{|0\rangle,|1\rangle\}$ of the Hilbert space $\sH$ we define an equivalence relation between state vectors: $|\psi\rangle\sim|\phi\rangle$ if and only if they differ by \emph{relative phases} w.r.t. the basis $\{|0\rangle,|1\rangle\}$ then they are physically indistinguishable by means of a measurement performed w.r.t. this basis.
\\
Let $|\psi\rangle=U_0(t)U_I(t)|0\rangle$ for some $t\geq 0$:
$$
U_0(t)U_I^*(t)|\psi\rangle\sim U_0^*(t)U_I^*(t)|\psi\rangle,
$$
where we use the fact that  free evolution operator (and its adjoint) produce only relative phases because the fixed basis is made by eigenstates of free Hamiltonian. 
\\
If we prove the following relation:
\beq\label{rel}
U_0^*(t)U_I^*(t)|\psi\rangle\sim U_I^*(t)U_0^*(t)|\psi\rangle=|0\rangle,
\eeq
then identity (\ref{id}) is also proved because it implies $U_0(t)U_I^*(t)|\psi\rangle=e^{i\theta}|0\rangle$ for some $\theta\in\bR$ depending on the operator $U_I(t)$.\\
Consider decomposition $U_I^*(t)|0\rangle=\alpha|0\rangle+\beta|1\rangle$, then:
$$U_0^*(t)U_I^*(t)|0\rangle=\alpha e^{iE_0t}|0\rangle+\beta e^{i E_1 t}|1\rangle,$$
$$U_I^*(t)U_0^*(t)|0\rangle=\alpha e^{iE_0t}|0\rangle+\beta e^{i E_0 t}|1\rangle,$$\\
that is $U_0^*(t)U_I^*(t)|0\rangle\sim U_I^*(t)U_0^*(t)|0\rangle$. The same argument implies $U_0^*(t)U_I^*(t)|1\rangle\sim U_I^*(t)U_0^*(t)|1\rangle$ thus relation (\ref{rel}) is true by linearity.

\vspace{0cm}

\section*{Acknowledgements}

This work is partially supported by CryptoLabTN.\\
 I am grateful to D. D'Alessandro for reading a first draft of the present work. I thank V. Moretti for some suggestions to improve the paper last version.

\end{document}